\documentclass[sigconf,nonacm]{acmart}

\usepackage{amsmath}
\usepackage{tcolorbox}
\usepackage{tabularx}
\usepackage{booktabs} 
\usepackage{graphicx}
\usepackage{geometry}
\usepackage{longtable}
\usepackage{adjustbox}
\usepackage{threeparttable}
\usepackage{soul}

\usepackage{enumitem}
\setitemize{noitemsep,topsep=0pt,parsep=0pt,partopsep=0pt,leftmargin=15pt}

\begin{document}
\title{Exploring Scientific Debt: Harnessing AI for SATD Identification in Scientific Software}

\author{Eric L. Melin}
    \affiliation{%
    \institution{Department of Computer Science}
    \institution{Boise State University}
    \city{Boise}
    \state{Idaho}
    \country{USA}}
    \email{ericmelin@u.boisestate.edu}

\author{Ahmed Musa Awon}
    \affiliation{%
    \institution{Department of Computer Science}
    \institution{University of Victoria}
    \city{Victoria}
    \state{BC}
    \country{Canada}}
    \email{ahmedmusa@uvic.ca}
    
\author{Nasir U. Eisty}
\affiliation{%
    \institution{Department of EECS }
	   \institution{University of Tennessee}
	   \city{Knoxville}
	   \state{TN}
	   \country{USA}
}
\email{neisty@utk.edu}

\author{Neil A. Ernst}
    \affiliation{%
    \institution{Department of Computer Science}
    \institution{University of Victoria}
    \city{Victoria}
    \state{BC}
    \country{Canada}}
    \email{nernst@uvic.ca}

\author{Shurui Zhou}
    \affiliation{%
    \institution{Department of ECE}
    \institution{University of Toronto}
    \city{Toronto}
    \state{Ontario}
    \country{Canada}}
    \email{shuruiz@ece.utoronto.ca}    

    \begin{abstract}
 \textit{\textbf{Context:}}  
Developers often leave behind clues in their code, admitting where it falls short, known as Self-Admitted Technical Debt (SATD). In the world of Scientific Software (SSW), where innovation moves fast and collaboration is key, such debt is not just common but deeply impactful. As research relies on accurate and reproducible results, accumulating SATD can threaten the very foundations of scientific discovery. Yet, despite its significance, the relationship between SATD and SSW remains largely unexplored, leaving a crucial gap in understanding how to manage SATD in this critical domain.  
\textit{\textbf{Objective:}}  
This study explores SATD in SSW repositories, comparing SATD in scientific vs. general-purpose open-source software and evaluating transformer-based models for SATD identification.  
\textit{\textbf{Method:}}  
We analyzed SATD in 27 scientific and general-purpose repositories across multiple domains and languages. We fine-tuned and compared 10 transformer-based models (100M–7B parameters) on 67,066 labeled code comments.  
\textit{\textbf{Results:}}  
SSW contains 9.25x more Scientific Debt and 4.93x more SATD than general-purpose software due to complex computations, domain constraints, and evolving research needs. Furthermore, our best model outperforms existing ones.  
\textit{\textbf{Conclusions:}}  
This study uncovers how SATD in SSW differs from general software, revealing its impact on quality and scientific validity. By recognizing these challenges, developers and researchers can adopt smarter strategies to manage debt and safeguard the integrity of scientific discovery.
\end{abstract}

\keywords{SATD Classification, AI Driven, Scientific Software, Scientific Debt}

\maketitle

\section{Introduction}
\label{sec:introduction}

Scientific software (SSW) has become indispensable in modern scientific research, serving as the backbone for a broad array of disciplines ranging from climate modeling and molecular biology to astrophysics and high-energy physics~\cite{segal2008developing, arvanitou2021software}. It facilitates formulating, testing, and validating scientific hypotheses by enabling complex calculations, data analysis, and simulations that would otherwise be impossible or prohibitively time-consuming to perform manually~\cite{HEATON2015207}. The significance of SSW lies not only in its ability to process massive datasets or simulate intricate phenomena but also in its direct influence on the validity of scientific results.
As such, errors or inefficiencies in SSW can potentially compromise the credibility of entire research fields, a concern that the scientific community can ill afford to ignore ~\cite{Miller2006ASN}.

Scientists often develop their approaches to software development, shaped by the need to rapidly respond to evolving research questions and immediate demands. Kelly~\cite{Kelly2013IndustrialSS} refers to this as an ``amethodical'' approach, where the urgency of research outcomes overshadows the need for structured, future-proof software design. The pressure to meet tight grant deadlines, publish results quickly, work within budget constraints, adapt to evolving research methodologies, and sometimes simply a lack of concern for long-term sustainability further amplifies this short-term focus. Hannay et al. \cite{Hannay2009HowDS} and Pinto et al. \cite{Pinto2018HowDS} support this view, showing that scientists frequently resort to ad-hoc practices, motivated by the pressing demands of their work rather than by a deliberate focus on long-term software maintainability.

However, the long-term maintenance and evolution of SSW may face challenges described by the metaphor of Technical Debt (TD), introduced by Cunningham in 1992~\cite{Cunningham1992TheWP}. TD refers to the internal tasks you choose not to perform now but that run the risk of causing future problems if they are not addressed. These deferred tasks, whether related to optimizing code, refining design, or improving documentation, accumulate as ``debt'' over time ~\cite{behutiye2017analyzing}. Like financial debt, TD incurs ``interest'' in the form of inefficiencies, errors, increased maintenance costs~\cite{tom2013exploration}, and, in the context of SSW, the risk of undermining the accuracy and credibility of scientific outcomes.

The consequences of neglecting this ``debt'' are not merely theoretical. The critical importance of SSW is highlighted by past incidents where undetected software faults led to severe scientific consequences. For example, in 2006, errors in scientific code contributed to the retraction of five high-profile scientific papers and the abandonment of numerous projects dependent on these findings~\cite{Miller2006ASN}. This incident exposed a pressing issue within the scientific community: the ``interest'' paid due to TD in SSW can result in severe setbacks in both credibility and scientific progress.

One way to gain deeper insight into the challenges mentioned above and other unknown factors affecting SSW development is through the lens of Self-Admitted Technical Debt (SATD). SATD refers to instances where developers explicitly acknowledge TD within the code, often through comments, commit messages, or documentation. These acknowledgments, typically in task annotations like TODO or FIXME, signal areas in the code that require further attention, such as incomplete features, temporary fixes, or refactoring~\cite{Storey2008Todo}. For example, \textit{``TODO - Move the next two subroutines to a new module called glad\_setup? This would be analogous to the organization of Glide.''}. Potdar and Shihab~\cite{Potdar2014} formalized SATD and identified its frequent occurrence as a byproduct of time pressure and temporary solutions in complex codebases.

While SATD can be a valuable means for understanding the state of a software, manually identifying and categorizing it is an extremely time-consuming process. For example, Maldonado et al. \cite{Maldonado2017UsingNL} reported spending 185 hours classifying 62,556 comments, demonstrating the significant effort required for manual SATD identification. This labor-intensive nature makes it clear that automated methods are needed to efficiently detect and classify SATD, reducing the burden of manual review while still providing valuable insights into software quality.

However, there’s a notable limitation: the majority of studies have focused on general-purpose open-source software, using datasets from popular languages like Java and Python. \textbf{The landscape of SATD detection remains largely unexplored when it comes to specialized fields like SSW, where scientific domain-specific knowledge plays a crucial role}. In SSW, SATD may manifest differently due to the complexity of the domain and the need for deep, specialized knowledge. This makes it critical to develop tailored SATD detection tools that can account for the unique challenges posed by scientific contexts.

A recent master's thesis~\cite{awon2024self} based on the examination of 28,680 code comments from 9 SSW repositories defined a novel SATD called \textbf{Scientific Debt} as \textit{the accumulation of sub-optimal scientific practices, assumptions, and inaccuracies within SSW that potentially compromise the validity, accuracy, and reliability of scientific results}, which is particularly relevant in the context of SSW.
While SSW is responsible for powering some of the most critical research in fields like astronomy, biology, and physics, its codebase is often overlooked in SATD studies. This thesis provides the first dataset capturing the nuances of code comments in SSW, with a focus on what the thesis calls ``Scientific Debt''. While it provides an initial dataset capturing the nuances of code comments in SSW, it does not offer a way to automatically detect this Scientific Debt, leaving researchers and practitioners without tools to proactively mitigate its long-term consequences which could undermine the credibility of entire fields of inquiry and impose escalating costs on future scientific endeavors.
Our study steps into that gap, presenting a fine-tuned transformer model approach, testing multiple preprocessing methods, hyperparameters, and 10 transformer-based models to detect, and categorize both well-known SATD categories~\cite{potdar2014exploratory} and Scientific Debt in SSW and open-source general-purpose software. These advanced transformer-based models allow for a comprehensive analysis of the SATD present in SSW, providing a more precise understanding of its nature.

This research addresses the gap in the literature on SATD in SSW by producing new insights.
Through data augmentation and hyperparameter testing during transformer fine-tuning, our model outperforms the best existing benchmarks. 
We classified 101k SATD instances in SSW and 96k SATD instances in general-purpose open-source software, revealing that SSW exhibits a unique form of SATD compared to general-purpose open-source software, marked by a significant presence of Scientific Debt. Furthermore, we observe that SSW contains multitudes more SATD than general-purpose open-source software, suggesting a lack of focus on SATD management in SSW.

The key contributions and main findings of this study can be summarized as follows:

\begin{itemize}
    \item \textit{Identifying and classifying Scientific Debt in SSW, providing the first formal analysis of this unique form of SATD.}
    \item \textit{Measuring and analyzing the differences in SATD between scientific and open-source software.}
\end{itemize}

\section{Background and Related Work}
\label{sec:background}

\subsection{Self-Admitted Technical Debt}
\label{subsec: SATD definitions}

Building on the foundational work of Potdar and Shihab~\cite{Potdar2014}, which introduced the concept of SATD, Maldonado and Shihab~\cite{maldonado2015detecting} advanced the field by categorizing SATD into distinct types. In their seminal 2015 study, they analyzed source code comments from five open-source Java projects and identified five specific SATD types: design debt (comments that indicate that there is a problem with the design of the code), defect debt (comments that state that a part of the code does not have the
expected behavior), documentation debt (comments that express that there is no proper documentation supporting that part of the program), requirement debt (comments that express incompleteness of the method, class or program), and test debt (comments that express the need for implementation or improvement of the current tests). This classification is critical because it reveals that not all TD is uniform. Each type reflects unique challenges and implications for software maintenance and evolution. Understanding these distinctions is essential for prioritizing remediation efforts, as, for instance, \emph{design debt} may signal deep architectural flaws requiring significant refactoring, while \emph{test debt} could indicate vulnerabilities in quality assurance that risk undetected defects. By differentiating SATD types, Maldonado et al. provide a framework that underscores why we should care: \emph{recognizing and addressing these categories enables developers and managers to mitigate specific risks, optimize resource allocation, and enhance long-term software sustainability}.

While these annotations are useful for tracking unresolved issues, they often remain unaddressed across multiple releases, leading to long-term maintenance challenges. Potdar and Shihab~\cite{Potdar2014} found that only 26.3\% to 63.5\% of SATD was resolved over time, and they identified 62 recurring patterns that help categorize these debts~\cite{potdar2014exploratory}. SATD encompasses various types, reflecting the different areas where TD can accumulate. For example, Liu et al.~\cite{Liu2021} conducted a study closely related to ours, focusing on SATD in deep learning frameworks, and found that design debt is the most frequently introduced type, followed by requirement and algorithm debt.

\subsection{Why Scientific Software Matters}
SSW development differs from traditional software practices due to its need for precision and longevity. Kelly et al.~\cite{Kelly2011SoftwareEF} emphasize that SSW handles large datasets and complex simulations, requiring both domain and computational expertise. However, scientists often lack formal software engineering training, relying on ad-hoc development methods~\cite{Wilson2006WheresTR}.
Wilson~\cite{Wilson2006WheresTR} notes that many scientists are self-taught, leading to unstructured code. Hannay et al.~\cite{Hannay2009HowDS} found that 96.9\% of scientists learn programming informally, and Pinto et al.~\cite{Pinto2018HowDS} observed similar trends in R developers. Both studies highlight challenges such as poor documentation and collaboration difficulties.
SSW is expected to be in use for decades, yet Kelly ~\cite{Kelly2009DeterminingFT} point out that evolving software with scientific and technological advancements is hard due to issues like inadequate documentation and SATD. Kelly and Sanders.~\cite{kelly2008assessing} and Segal~\cite{Segal2008ScientistsAS} note that scientists often prioritize research over software maintainability, leading to poorly optimized software.
Testing is also challenging. Carver et al.~\cite{Carver2007SoftwareDE} highlight the difficulty of ensuring correctness in complex scientific computations, and Kelly and Sanders~\cite{Sanders2008Challenge} identify the ``oracle problem" in validation. This, combined with limited testing knowledge, leads to unaddressed TD.
Kelly~\cite{Kelly2011AnAO} suggest hybrid development models to better address SSW needs. Despite proposed solutions, SATD in SSW remains unexplored, with many scientists unaware of its long-term impact~\cite{awon2024self}, emphasizing the need for automated SATD management tools.

\subsection{Automatic Identification of SATD}
Maldonado et al.~\cite{Maldonado2017UsingNL} pioneered the use of natural language processing (NLP) techniques to automatically detect SATD within source code comments. Their approach leveraged a maximum entropy classifier trained on features extracted from comments, facilitating the identification of common SATD types, such as design debt and requirement debt. This method demonstrated the potential of NLP in detecting SATD, highlighting how linguistic features in code comments can serve as indicators of TD.

Building on such foundations, Guo et al.~\cite{Guo2021HowFH} conducted a comprehensive review of SATD detection techniques, identifying key developments in both pattern-based and machine learning-based methods. Early approaches, such as those by Potdar and Shihab~\cite{potdar2014exploratory}, relied on manually defined comment patterns to detect SATD. While effective in certain contexts, this pattern-based method lacked scalability and precision in large projects. To overcome these limitations, machine learning techniques—such as text mining—were introduced. Huang et al.~\cite{Huang2017IdentifyingST} applied text mining to detect SATD across open-source projects, improving accuracy by training classifiers on labeled datasets. Guo et al.~\cite{Guo2021HowFH} also noted the growing use of convolutional neural networks (CNN) and recurrent neural networks (RNN) in SATD detection, which had shown promising results in enhancing detection accuracy in large codebases.

However, as automated approaches evolved, transformer-based models like BERT (Bidirectional Encoder Representations from Transformers) gained prominence for their ability to better understand the context of comments, thereby improving detection accuracy.
Yu et al.~\cite{Yu2022} introduced an advanced deep learning model for SATD detection, utilizing BERT to capture contextual information within comments. Their model significantly outperformed traditional NLP approaches by understanding the underlying semantics of the text, making it particularly effective in identifying SATD in complex projects. This approach demonstrated the potential of using transformer-based models to address the limitations of prior techniques, such as pattern-based detection.

More recently, with the widespread use of Large Language Models (LLMs), few studies have experimented with the effectiveness of utilizing LLMs for SATD identification and classification. Sheikhaei et al.~ \cite{sheikhaei2024empirical} evaluated the Flan-T5 family, outperforming the best existing non-LLM baseline with a best cross-project f1-score of 0.839. Lambert et al.~\cite{lambert2024identification} evaluated prompt engineering with Claude 3 Haiku, GPT 3.5 turbo, and Gemini 1.0 pro resulting in reduced LLM performance. Gu et al.~\cite{gu2024self} evaluated BERT on SATD identification across four different artifacts (code comments, issue tracker, pull requests, commit messages) reporting a cross-project f1-score of 0.859 on source code comments.
Furthermore, a recent systematic literature review from Melin and Eisty~\cite{melin2024exploringadvancesusingmachine} observed that transformer-based models outperformed traditional machine learning techniques for SATD identification and classification. 

Unlike prior works which all cover SATD in general-purpose open-source software, our study introduces the first approach explicitly tailored for \textbf{SATD analysis in SSW}, a critical yet understudied domain. In this paper, we investigate Scientific Debt—an unstudied type of SATD unique to scientific repositories—while presenting a new labeled augmented dataset, and a fine-tuned transformer model. 
Additionally, our work pioneers the evaluation of large-scale LLMs (exceeding 3B parameters) for SATD identification, with our best fine-tuned model \textbf{surpassing the state-of-the-art} cross-project F1-score reported by Gu et al.~\cite{gu2024self} by 0.0618 demonstrating superior performance. By leveraging fine-tuned transformers-based models, we tackle the distinct challenges of detecting and classifying SATD in the complex, domain-specific context of scientific codebases. 
Our study not only boosts detection accuracy but also reveals unique debt patterns in SSW, laying the groundwork for more effective technical debt management in SSW development.

\section{Research Questions}
\label{sec:research_questions}

Our study investigates the distinctions in SATD between open-source SSW and general-purpose open-source software. Our best fine-tuned model resulting from RQ1 identifies instances of SATD by file name and line number and then classifies them into one of six support types.
We pose the following research questions (RQs) to drive our study.

\textbf{RQ1: \textit{What fine-tuned transformer model performs best for SATD identification and classification?}}

Identifying and classifying SATD accurately is crucial for understanding its prevalence and impact across different software domains, such as SSW and general-purpose open-source software, as we will explore in RQ2. However, the effectiveness of SATD detection hinges on the choice of tools and methods, particularly with the advent of transformer-based models, which have revolutionized NLP tasks. This RQ addresses this foundational need by investigating which update-to-date fine-tuned transformer architecture performs better in extracting and categorizing SATD from code comments. By establishing the most effective transformer model for SATD identification and classification, RQ1 lays the groundwork for reliable, automated analysis, enabling us to confidently provide the most accurate results to RQ2.

\noindent \textbf{RQ2: \textit{Does scientific software contain different types and \newline amounts of SATD than general-purpose software?}}

Exploring the differences in SATD between open-source SSW and general-purpose open-source software can significantly advance our understanding of TD management in domain-specific contexts.
As discussed in Section~\ref{sec:background} while SATD in general-purpose software has been widely studied, the specific challenges of SATD in SSW remain underexplored.
Moreover, SSW development, often led by domain experts without formal software engineering training~\cite{carver2007software, Pinto2018HowDS, hannay2009scientists}, tends to accumulate more TD due to the need to translate complex scientific theories into computational models.
Therefore, we aim to compare SATD in SSW versus general-purpose open-source software, shedding light on which types of SATD are more prevalent in SSW.

\section{Methodology}
\begin{figure}[h]
    \centering
    \includegraphics[width=1\linewidth]{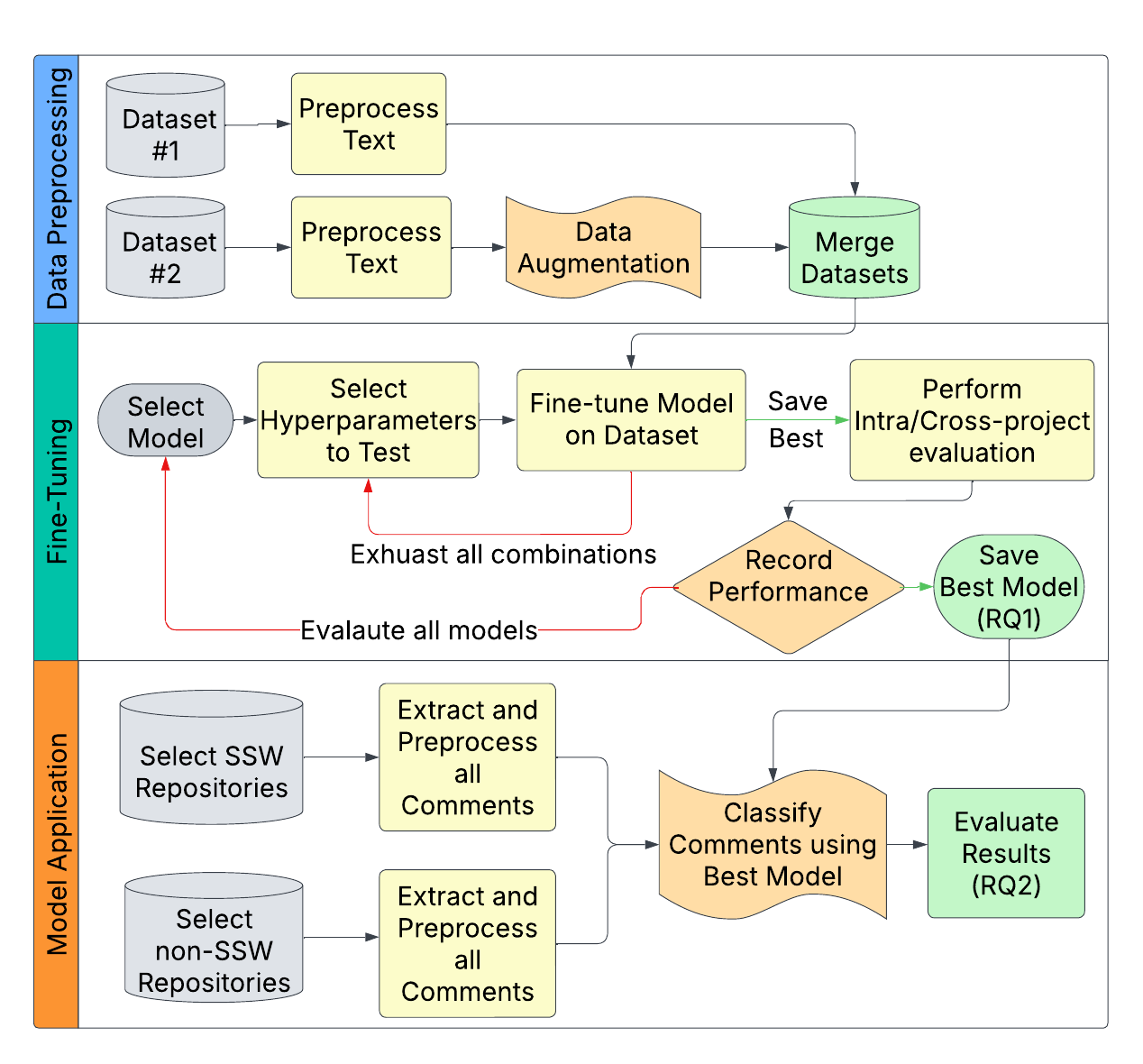}
    \caption{Approach Overview}
    \label{fig:Approach Overview}
\end{figure}

To answer the proposed RQs, we first perform data preprocessing where we preprocess, augment, and merge two existing labeled datasets to prepare for model fine-tuning discussed in Section \ref{subsec:datasets}.
We then conduct throughout hyperparameter testing on selected transformer models to maximize intra-project and cross-project performance of our models on our merged dataset discussed in Section \ref{subsec:fine-tuning}.
Finally, we extract and preprocess source code comments from unlabeled open-source SSW and general-purpose open-source software repositories and apply the best performing model to identify, classify, and locate SATD instances across the repositories as described in Section~\ref{subsec: Model Application}. An overview of our approach is illustrated in \textbf{Figure~\ref{fig:Approach Overview}}.

\label{sec:methodology}

\subsection{Dataset}
\label{subsec:datasets}
To observe if SSW features differences from general-purpose open-source software and fine-tune a model to automatically identify and categorize SATD, we first needed to combine a labeled dataset featuring traditional forms of SATD with the dataset containing labeled instances of Scientific Debt.
For this study, the merge of the two selected datasets results in 67k labeled SATD comments from across 19 repositories.

\subsubsection{First Dataset:} 
We used the SATDAUG dataset by Sutoyo and Capiluppi~\cite{sutoyo2024satdaug}, comprising 69k labeled source code comments.
SATDAUG augments manually labeled SATD comments from minority classes to balance the dataset across supported SATD types, addressing class imbalance issues to enhance detection accuracy in prior work~\cite{sridharan2021data}.
The original comments, pre-augmentation, originate from 10 general-purpose open-source Java projects~\cite{li2023automatic}.

SATDAUG supports four SATD types, defined in Section \ref{subsec: SATD definitions}:
\begin{itemize}
  \item  Code/Design Debt (C/D)
  \item Documentation Debt (DOC)
  \item Test Debt (TES) 
  \item  Requirement Debt (REQ)
\end{itemize}

Foundational research for SATDAUG~\cite{da2017using} observed that code and design debt are not distinctly separable in comments due to their overlap, leading SATDAUG to combine them as Code/Design Debt.
The dataset in total includes 58k Non-SATD source code comments and 
2k labeled comments for the remaining four SATD types.
Furthermore, SATDAUG includes balanced data from multiple sources including: source code comments, issue trackers, pull requests, and commit messages. For this study, we focus solely on the source code comment subset (C/D, DOC, TES, REQ, and Non-SATD), aligning with our second dataset’s (Section \ref{subsec: Second Dataset}) exclusive use of comment-based labels.

\subsubsection{Second Dataset:}
\label{subsec: Second Dataset}
The second labeled dataset leveraged in this study originates from a recently published master's thesis \cite{awon2024self}. This work highlights a key distinction between traditional forms of SATD, which primarily pertain to software engineering concerns such as code quality and design practices, and the challenges associated with SSW. Specifically, the author observes that SSW faces unique difficulties in translating complex scientific methodologies into computational models. To capture these challenges, they introduce a novel SATD category termed \textbf{Scientific Debt}, defined as \textit{the accumulation of suboptimal scientific practices, assumptions, and inaccuracies within SSW that potentially compromise the validity, accuracy, and reliability of scientific results.}  

As part of their study, the authors compiled a dataset comprising 1,301 labeled comments indicative of Scientific Debt. For instance, a comment from the Astropy~\cite{robitaille2013astropy} project states:  
\begin{quote}
\emph{We are going to share neff between the neutrinos equally. In detail, this is not correct, but it is a standard assumption because properly calculating it is (a) complicated (b) depends on the details of the massive neutrinos.}  
\end{quote}  
Similarly, within the Elmer project, another comment reads~\cite{awon2024self}:  
\begin{quote}
\emph{For some reason it doesn't appear to give convergence... Find a remedy!}  
\end{quote}  
These examples illustrate the types of challenges documented within the dataset, reflecting the inherent complexities of maintaining scientific rigor in computational implementations, which have not yet been validated in SATD research.

To this end, we include this novel SATD in our experiments to explore if SSW contains a distinct SATD type and evaluate the differences between the two domains of software.

\subsubsection{Merged Dataset:}
The final dataset we used in this study is a combination of the two previously mentioned datasets utilized to answer RQ1 and RQ2. 
After applying the preprocessing steps listed below, from the first dataset we inherit: 2,703 C/D comments, 2,701 DOC comments, 2,635 TES comments, 2,269 REQ comments, and 54,237 Non-Debt comments all pertaining to only source code comments. 
From the SSW Dataset, there were 1,301 Scientific Debt comments. To verify the claims from~\cite{sutoyo2024satdaug, sridharan2021data} that data balancing improves SATD detection, and to provide better balance across our minority classes, we utilized AugGPT~\cite{dai2025auggpt}, a text augmentation technique based on the ChatGPT language model, to double the number of Scientific Debt instances available. Our objective was to generate an additional paraphrased version of each existing text while preserving the original meaning, since disregarding the original semantics of the text could result in mislabeling of the labeled class~\cite{xie2022multi}. 

\begin{figure}[h]
    \centering
    \fbox{\includegraphics[width=1\linewidth]{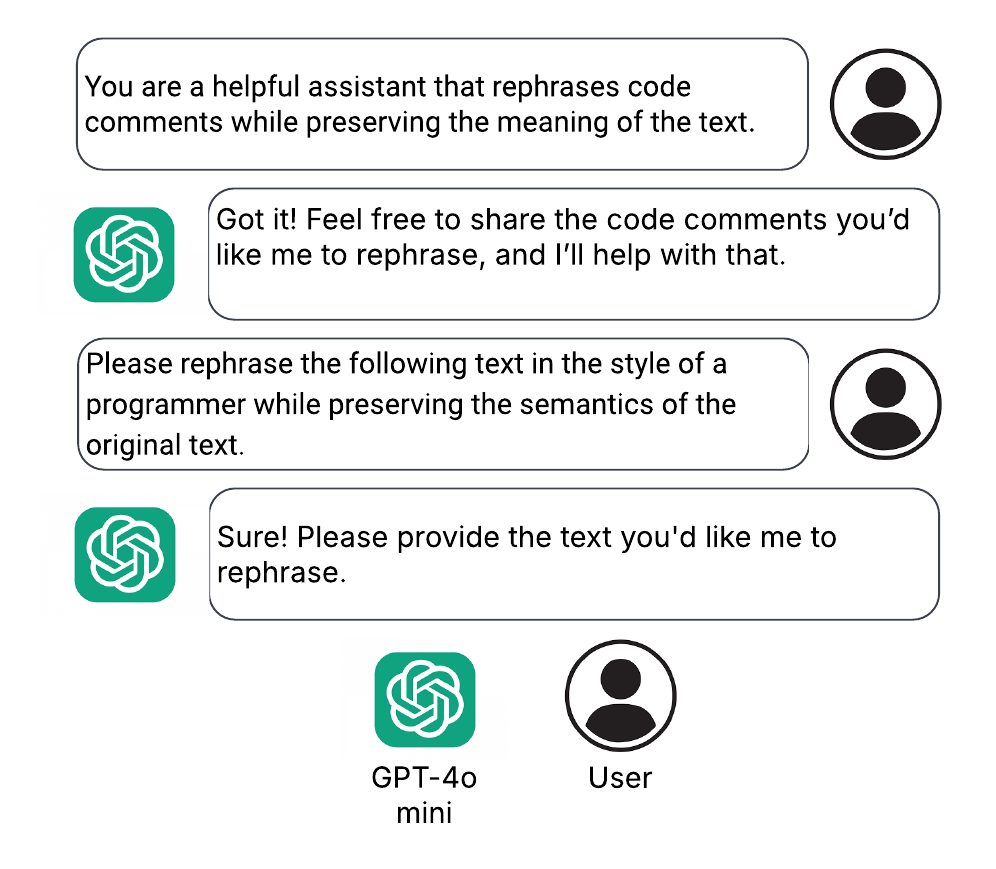}}
    \caption{Multi-turn dialog}
    \label{fig:Multi-turn dialog}
\end{figure}

To do this, we adopted a multi-turn prompt dialog approach leveraged on OpenAI's GPT-4o mini, guided by the principles outlined in Dai et al.~\cite{dai2025auggpt}. An example of our multi-turn prompt dialog for augmenting Scientific Debt comments used in this study can be seen in \textbf{Figure ~\ref{fig:Multi-turn dialog}}. The result of augmenting our Scientific Debt labels to provide balance with the remaining minority classes provided an increase to our intra-project F1-score by  0.68\%. This increase is the result of balancing our last minority class to contain a similar amount of labeled comments across all labeled SATD types.

Following Scientific Debt comment augmentation, we applied text preprocessing steps to our unprocessed dataset. First, we convert comments stretching multiple lines into a single line comment. Then, because different developers have different writing habits for SATD, for example, they may write ``TODO'' as ``Todo'', ``todo:'', etc., we refer to the preprocessing procedures of Huang et. al~\cite{huang2018identifying} and Maldonado et al.~\cite{da2017using}: (1) \textbf{Remove all non-alphabetic characters from the samples}, this removes code comment delimiters and miscellaneous symbols. Note that quotation and exclamation marks are preserved as they are deemed helpful for identifying SATDs by Maldonado et al.~\cite{da2017using}. (2) \textbf{Convert all text to lowercase}. Additionally, we stripped extra whitespace from comments, and removed empty lines.

Seminal SATD research utilizing text-mining and NLP techniques conducted experiments on the effect of stop word inclusion for SATD identification. To extend on this, we experiment on the inclusion and exclusion of stop words lists on transformer-based models. We tested the stop word list provided by Huang et al.~\cite{huang2018identifying} who for their study created a stop word list for SATD identification using text mining, the NLTK~\cite{https://doi.org/10.48550/arxiv.cs/0205028} stop words list, and omission of stop word removal to use as a baseline. The results show that our baseline of omitting stop word removal provided the best results on our transformer models, with the Huang et al.~\cite{huang2018identifying} list lowering the F1-score by 0.0043\% and the NLTK list lowering F1 by 0.0068\%.

After merging and preprocessing the two datasets, we find that from the 67,066 labeled comments, the average length of a comment is 9.70 words with a standard deviation of 19.52 words, which we take into account for selecting a max text length used for fine-tuning our transformer-based models. Table~\ref{tab:training_dataset} shows the counts of classified comments used in the training dataset for this study. 

\begin{table}[h!]
\normalsize
\begin{tabular}{lc}
\hline
\textbf{Comment Label} & \textbf{Count} \\ \hline
Non-SATD     & 54,237      \\
Code/Design Debt     & 2,703 \\ 
Documentation Debt     & 2,701 \\
Test Debt     & 2,635     \\ 
Requirement Debt     & 2,269      \\ 
Scientific Debt     & 2,521      \\ \hline
\end{tabular}
\caption{Training Dataset}
\label{tab:training_dataset}
\end{table}

\subsection{Selected Models}
\label{subsec: Selected Models}
In this study, we focus solely on fine-tuning transformer-based language models for the task of SATD identification and classification. The following models were evaluated:

\begin{itemize}
\item \textbf{BERT-base} \cite{DBLP:journals/corr/abs-1810-04805}: A widely-used transformer model known for its strong performance on NLP tasks.
\item \textbf{BERT-large} \cite{DBLP:journals/corr/abs-1810-04805}:  A larger variant of BERT with increased model capacity for better contextual understanding.
\item \textbf{CodeBERT-base} \cite{feng2020codebert}: A transformer model specifically pre-trained on source code and natural language, making it suitable for software engineering tasks.
\item \textbf{RoBERTa} \cite{DBLP:journals/corr/abs-1907-11692}: A BERT variant optimized through additional pretraining and dynamic masking, improving performance on various classification tasks.
\item \textbf{Mistral-7B} \cite{jiang2023mistral7b}: A highly efficient transformer model designed for superior language modeling and sequence prediction.
\item \textbf{Llama-2-7B} \cite{touvron2023llama2openfoundation}: A 7-billion parameter model developed by Meta, demonstrating strong performance on text classification.
\item \textbf{Deepseek Qwen} \cite{deepseekai2025deepseekr1incentivizingreasoningcapability}: Evaluated at both 1.5B and 7B parameter sizes, these models leverage advanced tokenization and pretraining techniques for improved generalization.
\item \textbf{T5-base} \cite{2020t5}: A sequence-to-sequence transformer trained with a denoising objective, allowing flexibility in classification.
\item \textbf{T5-large} \cite{2020t5}: A larger variant of T5 with increased parameters, improving contextual understanding and classification ability.
\end{itemize}

For both intra-project evaluation and cross-project evaluation we retrieve the following tokenizers and models via HuggingFace transformers utilizing AutoTokenizer, and AutoModelForSequenceClassification for fine-tuning: "google-bert/bert-base-uncased" for BERT-base, "google-bert/bert-large-uncased" for BERT-large, "microsoft/ codebert-base" for CodeBERT-base, "FacebookAI/roberta-base" for RoBERTa, "mistralai/Mistral-7B-v0.3" for Mistral-7B, "meta-llama/ Llama-2-7b-hf" for Llama-2-7B, 
"deepseek-ai/DeepSeek-R1-Distill-Qwen-1.5B" for Deepseek Quen 1.5B, "deepseek-ai/DeepSeek-R1-Distill-Qwen-7B" for Deepseek Qwen 7B, "google-t5/t5-base" for T5-base, and "google-t5/t5-large" for T5-large. 

\subsection{Fine-Tuning}
\label{subsec:fine-tuning}
To fine-tune these models, we leverage our institution's HPC cluster. All evaluations were run using NVIDIA L40s each with 48GB GDDR6 memory. 

All models were subject to thorough hyperparameter testing to maximize performance. For each model, we evaluate the following: \textit{Learning Rates} $\in \{1e^{-5}, 5e^{-5}, 1e^{-4}\}$ and \textit{Weight Decays} $\in \{0.0, 0.01, 0.1\}$. For each model, we also select the largest batch size available, constrained by the selected model size on our GPUs, including \textit{Batch Sizes} $\in \{256, 128, 64, 32, 16, 8\}$, to smooth gradient variance and maximize training efficiency.

Given our labeled dataset of 67,066 labeled comments with an average length of 9.70 words and standard deviation of 19.52 words, we select a max length of 128 during training to allow for larger batch sizes while ensuring that 99\% of labeled comments are still available for fine-tuning. For intra-project validation, we split our shuffled dataset into 80\% train, 10\% validation, and 10\% test. 

Each corresponding model's tokenizer is then employed to tokenize textual inputs, ensuring uniform input length through padding and truncation to a maximum sequence length of 128 tokens. A custom PyTorch dataset was implemented to encode the textual data and map classification labels to numerical values using a predefined mapping scheme for each of the six labeled classes of text.

Each pre-trained model is then fine-tuned for sequence classification with six output classes, corresponding to the predefined classification labels. The model was trained on a GPU-accelerated environment with the maximum batch size.

Then our hyperparameter tuning process conducts over a grid search of our predefined learning rates and weight decays. The AdamW optimizer was used with a linear scheduler incorporating a warm-up phrase to stabilize training. 

The training process for models with less than 7B parameters spanned a maximum of 10 epochs with early stopping criterion, terminating if validation loss did not improve over two consecutive epochs to prevent overfitting. Cross-entropy loss was employed as the objective function.

To fine-tune the 7B parameter models efficiently on our GPU hardware, we employed Low-Rank Adaptation (LoRA) \cite{hu2021loralowrankadaptationlarge}. LoRA reduces the number of trainable parameters by decomposing the weight updates into low-rank matrices, significantly lowering memory requirements while maintaining fine-tuning effectiveness. The selected LoRA configuration was
$\{r=16, \text{lora\_alpha}=16, \newline \text{lora\_dropout}=0.1, \text{task\_type}=\text{TaskType.SEQ\_CLS}\}$. 
Each model was trained for a total of three epochs using the same early stopping criterion, with cross-entropy loss as the objective function.

\subsection{Data Preparation for Model Application}
\label{subsec: Model Application}
In order to answer RQ2 utilizing our best performing fine-tuned model from RQ1, we selected a total of 27 repositories unique from the training datasets, which we categorize into two distinct groups~\cite{harris2020array} to evaluate SATD in SSW and general-purpose open-source software. The selected repositories primarily utilize Perl, Fortran, C++, C, Python, Typescript, Go, or PHP.

\begin{itemize}

\item Domain-Level \& Foundation Scientific Repositories: We chose 16 repositories from various scientific domains, including Applied Mathematics, Climate Modeling, Astronomy, Physics, High Energy Physics, Molecular Dynamics, Molecular Biology, Machine Learning, etc.

\item General-purpose Open-Source Software Repositories: We chose eleven repositories to encompass a broad spectrum of open-source projects.
\end{itemize}

Criteria for Selection:
\begin{itemize}

\item Popularity: All repositories have a minimum of 40 GitHub stars.
\item Engagement: Each repository has at least 15 total contributors.
\item Activity: All repositories have been updated since 2023.
\end{itemize}

A general overview of the scientific repositories can be seen in Table~\ref{tab:Overview of Case Study Scientific Repositories}. To count the total lines of code (LOC) we used CLOC ~\cite{aldanial_2024_13179769}.

\begin{table*}
\normalsize 
\centering
\caption{Overview of Selected Scientific Repositories}
\begin{tabular}{llrrrrr}
\hline
\textbf{Name} & \textbf{Domain} & \textbf{Cont.} & \textbf{LOC} & \textbf{Stars} & \textbf{Forks} & \textbf{Age (Yrs)} \\ \hline
Bioperl-live & Molecular Biology      & 72    & 274k  & 294   & 181   & 14 \\ 
Healpy       & Physics        & 70    & 13k   & 255   & 186   & 13 \\ 
LAMMPS       & Molecular Dynamics      & 266   & 1923k & 2.2k  & 1.7k  & 11 \\ 
Mad-X        & High Enenergy Physics & 22    & 386k  & 48    & 39    & 7  \\ 
Matplotlib   & Data Visualization    & 1.5k & 606k  & 21k   & 7.8k  & 22 \\ 
MDAnalysis   & Molecular Dynamics      & 197  & 84k   & 1.3k  & 644   & 9  \\ 
MITgcm       & Climate Model    & 65    & 1153k & 325   & 240   & 6  \\ 
NumPy        & Applied Mathematics & 1.7k & 548k & 29k   & 10.5k  & 20 \\ 
OpenMM       & Molecular Dynamics      & 199   & 699k  & 1.5k  & 510   & 11 \\ 
Pandas       & Data Analysis      & 3.4k  & 551k  & 44.7k   & 18.3k  & 16 \\ 
PyTorch      & Machine Learning   & 3.7k & 2.5M & 87.5k   & 23.5k   & 9  \\ 
Scikit-learn & Data Analysis      & 2.9k  &  282k & 61.3k   & 25.6k   & 18 \\ 
SciPy        & Applied Mathematics & 1.6k & 657k  & 13.4k   & 5.3k  & 24 \\ 
SPLASH       & Astronomy      & 18    & 64k   & 56    & 41    & 7  \\ 
sympy        & Applied Mathematics    & 1.2k  & 477k  & 12.7k & 4.4k  & 14 \\ 
TensorFlow   & Machine Learning   & 3.6k & 4.4M & 188k  & 74.6k   & 10 \\ \hline
\end{tabular}

\label{tab:Overview of Case Study Scientific Repositories}
\end{table*}

\section{Results}
\label{sec:results}

In this section, we evaluate the results of fine-tuning multiple models and applying our best classifier to the selected repositories.  

\subsection{RQ1: What fine-tuned transformer model performs best for SATD identification and classification?}

Intra-project evaluation was conducted using the validation dataset to monitor the model’s learning progression, with final performance evaluated on the test set.
The performance metrics employed included precision, recall, macro F1-score, weighted F1-score, and support.
The optimal hyperparameter configuration was determined by selecting the model with the highest weighted F1-score.
\textbf{Table~\ref{tab:intra_model_performance}} presents the intra-project performance of the 10 selected models (detailed in Section \ref{subsec: Selected Models}), each with optimized hyperparameters, across the six SATD classes, alongside the macro-average F1-score and weighted-average F1-score.
The results indicate that all models exhibit comparable performance, with BERT-large achieving the highest intra-project SATD classification performance, evidenced by a weighted-average F1-score of 0.9827. To address RQ2, we utilize this fine-tuned BERT-large model, which was trained on data encompassing all labeled projects. 

\begin{table*}
    \centering
    \normalsize
    \caption{Intra-Project Model Evaluation}
    \begin{tabular}{l|cccccc|cc}
        \toprule
        Model & REQ F1& C/D F1& DOC F1& TES F1& SCI F1& Non-SATD F1& Macro Avg F1 & Weighted Avg F1 \\
        \midrule
        BERT-base  & 0.8894 & 0.8114 & 0.9981 & 0.9925 & 0.9533 & 0.9921 & 0.9395 & 0.9809 \\
        BERT-large & 0.8856 & \textbf{0.8409} & \textbf{1.0000} & 0.9944 & 0.9545 & 0.9929 & 0.9447 & \textbf{0.9827} \\
        CodeBERT-base  & 0.8806 & 0.7992 & 0.9962 & 0.9925 & 0.9622 & 0.9915 & 0.9371 & 0.9799\\
        RoBERTa  & 0.8641 & 0.7793 & \textbf{1.0000} & 0.9797 & 0.9509 & 0.9906 & 0.9274 & 0.9771 \\
        Mistral-7B  & 0.8678 & 0.8057 & 0.9962 & 0.9925 & 0.9662 & \textbf{0.9932} & 0.9370 & 0.9813 \\
        Llama-2-7B  & 0.8824 & 0.8144 & \textbf{1.0000} & \textbf{0.9963} & 0.9576 & 0.9923 & 0.9405 & 0.9813 \\
        Deepseek-Qwen 1.5B  & 0.8840 & 0.8171 & 0.9963 & 0.9944 & \textbf{0.9666} & 0.9930 & 0.9419 & 0.9821 \\
        Deepseek-Qwen 7B  & 0.8629 & 0.8016 & 0.9962 & 0.9925 & 0.9407 & 0.9914 & 0.9309 & 0.9786 \\
        T5-base  & 0.8634 & 0.8024 & 0.9963 & 0.9944 & 0.9317 & 0.9920 & 0.9300 & 0.9788 \\
        T5-large & \textbf{0.8950} & 0.8304 & \textbf{1.0000} & 0.9963 & 0.9609 & 0.9926 & \textbf{0.9459} & 0.9826 \\
        \bottomrule
    \end{tabular}
    \label{tab:intra_model_performance}
\end{table*}

Cross-project evaluation was performed leveraging the optimal hyperparameters from the previous search, on a stratified group k-fold of our labeled dataset to ensure a similar class balance across all folds. For this evaluation, we test on 5 folds, ensuring that the training data from one project can not be used for testing. Our stratified group 5-fold averaged 15 projects for training and 4 projects for testing for each fold. Differing from our intra-project evaluation, we select to train for 5 epochs for each fold, with a validation loss patience of 2 before termination. To fairly evaluate the models on each of the 5 folds, we report the averaged values across the 5 folds. The results from our averaged cross-project evaluation can be seen in Table \ref{tab:cross_model_performance}. Note that for cross-project evaluation for 7B models, we additionally leverage 8-bit quantization to manage memory constraints. Here we observe that for cross-project evaluation LLama-2-7B performs the best with an averaged macro avg f1-score of 0.7621 and averaged weighted avg f1-score of 0.9337.

\begin{table*}
    \centering
    \normalsize
    \caption{Averaged 5-Fold Cross-Project Model Evaluation}
    \begin{tabular}{l|cccccc|cc}
        \toprule
        Model & REQ F1& C/D F1& DOC F1& TES F1& SCI F1& Non-SATD F1& Macro Avg F1 & Weighted Avg F1 \\
        \midrule
        BERT-base  & 0.5774 & 0.5341 & 0.7855 & 0.7679 & 0.6654 & 0.9775 & 0.7189 & 0.9222 \\
        BERT-large & 0.5281 & 0.4865 & 0.7992 & 0.7272 & 0.6983 & 0.9763 & 0.7026 & 0.9208 \\
        CodeBERT-base  & 0.5548 & 0.5731 & 0.8083 & 0.7607 & 0.7066 & 0.9793 & 0.7305 & 0.9290\\
        RoBERTa  & 0.5688 & 0.5001 & 0.7649 & 0.7621 & 0.6706 & 0.9766 & 0.7072 & 0.9185 \\
        Mistral-7B  & 0.5919 & \textbf{0.5904} & 0.7891 & 0.8062 & 0.7381 & 0.9787 & 0.7491 & 0.9274 \\
        Llama-2-7B   & \textbf{0.5946} & 0.5858 & 0.8403 & 0.8066 & \textbf{0.7657} & \textbf{0.9797} & \textbf{0.7621} & \textbf{0.9337} \\
        Deepseek-Qwen 1.5B  & 0.5673 & 0.5299 & 0.6931 & 0.6189 & 0.6638 & 0.9753 & 0.6747 & 0.9118 \\
        Deepseek-Qwen 7B  & 0.5176 & 0.5477 & 0.8138 & 0.8115 & 0.7405 & 0.9729 & 0.7340 & 0.9206 \\
        T5-base  & 0.5582 & 0.5334 & 0.7682 & \textbf{0.8309} & 0.6512 & 0.9787 & 0.7201 & 0.9200 \\
        T5-large & 0.5867 & 0.5502 & \textbf{0.8415} & 0.7803 & 0.7102 & 0.9770 & 0.7410 & 0.9251 \\
        \bottomrule
    \end{tabular}
    \label{tab:cross_model_performance}
\end{table*}

\begin{tcolorbox}[
    title=Summary of RQ1 Results,  
    colback=blue!5!white,        
    colframe=blue!75!black,      
    coltitle=white,              
    fonttitle=\bfseries,         
    colbacktitle=blue!50!black,  
    boxrule=0.75mm,              
    rounded corners              
]
Our thorough analysis of fine-tuning transformers displayed impressive performance across all selected models. We observe that for intra-project evaluation BERT-large narrowly outperforms large models for weighted f1-score, while for cross-project evaluation we observe that the 7B models display the best performance with Llama-2-7B performing the best for both macro and weighted f1-scores.
\end{tcolorbox}

\subsection{RQ2: Does scientific software contain different types and amounts of SATD than general-purpose software?}

To analyze if there are differences between SSW and general-purpose open-source software, we applied our best fine-tuned performing transformer model BERT-large, to all the selected scientific and general-purpose repositories. 
We selected 16 scientific and 11 general-purpose open-source repositories constrained to our repository selection criteria to answer RQ2. Tables ~\ref{tab:Scientific Repositories Results} and Table~\ref{tab:Open Source Repositories Results} show tabulated results of running fine-tuned BERT-large on the selected repositories. 

\begin{table*}
\normalsize
\centering
\caption{Scientific Repositories Results. Note: * Implies Column Average. }
\label{tab:Scientific Repositories Results}
\begin{threeparttable}
\begin{tabular}{p{2.5cm}p{2.8cm}p{.8cm}p{1.0cm}rrrrrrr}
\hline
\textbf{Repo Name} & \textbf{Repo Domain} & \textbf{Total SATD} & \textbf{Non SATD} & \textbf{DOC} & \textbf{REQ} & \textbf{TES} & \textbf{C/D} & \textbf{SCI} & \textbf{\% SCI} & \textbf{\% SATD} \\
\hline
Bioperl-live & Molecular Biology & 1,192 & 6,891 & 9 & 17 & 7 & 158 & 1,001 & 12.38 & 14.75 \\
Healpy & Physics & 177 & 367 & 0 & 1 & 1 & 4 & 171 & 31.43 & 32.54 \\
LAMMPS & Molecular Dynamics & 19,022 & 71,485 & 24 & 153 & 124 & 577 & 18,144 & 20.05 & 21.02 \\
Mad-X & High Energy Physics & 2,559 & 21,589 & 3 & 51 & 20 & 186 & 2,299 & 9.52 & 10.6 \\
Matplotlib & Data Visualization & 1,810 & 14,584 & 20 & 50 & 46 &138 & 1,556 & 9.49 & 11.04 \\
MDAnalysis & Molecular Dynamics & 2,125 & 4,764 & 17 & 6 & 33 & 53 & 2,016 & 29.26 & 30.85 \\
MITgcm & Climate Model & 318 & 2,558 & 15 & 4 & 1 & 29 & 269 & 10.52 & 12.43 \\
Numpy & Applied Mathematics & 4,546 & 30,616 & 13 & 104 & 83 & 315 & 4,031 & 11.46 & 12.93 \\
OpenMM & Molecular Dynamics & 5,263 & 17,458 & 8 & 56 & 45 & 67 & 5,087 & 22.39 & 23.16 \\
Pandas & Data Analysis & 2,914 & 28,619 & 16 & 119 & 229 & 506 & 2,044 & 6.48 & 9.24 \\
Pytorch & Machine Learning & 20,238 & 89,267 & 62 & 935 & 782 & 2,511 & 15,948 & 14.56 & 18.48 \\
Scikit-learn & Data Analysis & 5,519 & 15,512 & 4 & 41 & 145 & 211 & 5,118 & 24.34 & 26.24 \\
SciPy & Applied Mathematics & 7,981 & 24,593 & 23 & 58 & 115 & 112 & 7,673 & 23.56 & 24.50 \\
SPLASH & Astronomy & 1040 & 4,162 & 0 & 4 & 0 & 12 & 1,024 & 19.68 & 19.99 \\
sympy & Applied Mathematics & 5,674 & 19,804 & 21 & 83 & 152 & 317 & 5,101 & 20.02 & 22.27 \\
Tensorflow & Machine Learning & 20,357 & 87,497 & 51 & 913 & 1,789 & 1,774 & 15,830 & 14.68 & 18.87 \\
\hline
\textbf{Total} & & 100,735 & 439,766 & 286 & 2,595 & 3,572 & 6,970 & 87,312 & 17.49* & 19.31* \\
\hline
\end{tabular}
{Note: \textbf{DOC}: Documentation Debt; \textbf{REQ}: Requirement Debt; \textbf{TES}: Test Debt; \textbf{C/D}: Code/Design Debt;  \textbf{SCI}: Scientific Debt.}
\end{threeparttable}
\end{table*}

\begin{table*}
\small
\centering
\caption{Open Source Repositories Results. Note: * Implies Column Average. }
\label{tab:Open Source Repositories Results}
\begin{tabular}{p{2.9cm}p{3.0cm}p{.8cm}p{1.0cm}rrrrrrr}
\hline
\textbf{Repo Name} & \textbf{Repo Domain} & \textbf{Total SATD} & \textbf{Non SATD} & \textbf{DOC} & \textbf{REQ} & \textbf{TES} & \textbf{C/D} & \textbf{SCI} & \textbf{\% SCI} & \textbf{\% SATD} \\
\hline
Caddy & Web Server/Infrastructure & 199 & 4,858 & 4 & 29 & 30 & 77 & 59 & 1.17 & 3.94 \\
Django-rest-framework & Web Development & 40 & 1,934 & 1 & 3 & 14 & 11 & 11 & 0.56 & 2.03 \\
GIMP & Image Editing & 1,447 & 25,838 & 13 & 111 & 22 & 484 & 817 & 2.99 & 5.30 \\
Home-Assistant-core & Smart Home & 5,597 & 136,848 & 19 & 324 & 940 & 152 & 4,162 & 2.92 & 3.93 \\
Hugo & Static Site Generator & 243 & 8,411 & 7 & 20 & 32 & 113 & 71 & 0.82 & 2.81 \\
Kubernetes & Container Orchestration & 10,066 & 208,669 & 252 & 1,474 & 1,375 & 2,095 & 4,870 & 2.23 & 4.60 \\
Moby & Containerization & 5,223 & 105,347 & 117 & 809 & 496 & 1,169 & 2,632 & 2.38 & 4.72 \\
Mozilla Firefox & Web Browsing & 68,585 & 1,122,068 & 2,562 & 5,156 & 6,442 & 14,339 & 40,086 & 3.37 & 5.76 \\
Next-Cloud-server & Cloud Storage & 666 & 47,395 & 6 & 124 & 57 & 374 & 105 & 0.22 & 1.39 \\
VLC & Media Playback & 2,233 & 34,431 & 22 & 226 & 75 & 668 & 1,242 & 3.39 & 6.09 \\
VSCode & Software Development & 1,573 & 60,947 & 28 & 252 & 65 & 773 & 455 & 0.75 & 2.58 \\
\hline
\textbf{Total} & & 95,872 & 1,756,744 & 3,031 & 8,528 & 9,548 & 20,255 & 54,510 & 1.89* & 3.92* \\
\hline
\end{tabular}
\end{table*}

To evaluate whether scientific repositories differ from general-purpose open-source repositories, we leveraged our fine-tuned BERT-large model to classify all comments in each GitHub repository. Then, we compared the percent of comments classified as Scientific Debt comments to all comments classified, since all the repositories vary in size. This percent for each repository can be seen in the \textit{\% SCI} column of each table. 
The analysis of this column reveals that Scientific Debt is highly prevalent in scientific repositories, at an average of 17.49\% of all comments in each repository. In contrast, general-purpose open-source repositories have an average of 1.89\% of total comments classified as Scientific Debt. 
These results suggest a difference between the nature of SATD between SSW and general-purpose open-source software. Since we see only high amounts of Scientific Debt in SSW and low amounts in general-purpose open-source software, we see that SSW contains a unique type of SATD and that our model works well on various repositories.

Furthermore, when analyzing the overall distribution of classified comments across the two types of software, we observe a trend of SSW containing significantly more SATD than general-purpose open-source software. The percent of total SATD in each repository can be observed in the \textit{\% SATD} column of each table. We observe that the classified SSW repositories contain an average of 19.31\% of all comments indicating SATD whereas general-purpose open-source software on average contains a mere 3.92\% SATD. 
This large difference in SATD prevalence between the two types of repositories suggests a need for SSW specific SATD management and a need for tools catered specifically to SSW since it exhibits differences in SATD from general-purpose open-source software (discussed in section~\ref{sec:Research Question Analysis and Implications}).

When observing the distribution of traditional types of SATD between the two domains of software, we also observe differences. For instance, comparing the total SATD instances by type to the total comments analyzed in each domain, we observe that general-purpose open-source software contains 3.09x more documentation debt while SSW contains 1.04x more requirement debt, 1.28x test debt, and 1.18x code/design debt.
These findings could suggest that while SSW contains more total SATD, SSW performs especially well regarding documentation debt, likely due to the nature of scientific procedures requiring thorough documentation.

\begin{tcolorbox}[
    title=Summary of RQ2 Results,  
    colback=blue!5!white,        
    colframe=blue!75!black,      
    coltitle=white,              
    fonttitle=\bfseries,         
    colbacktitle=blue!50!black,  
    boxrule=0.75mm,              
    rounded corners              
]
Our comparison of SATD between SSW and general-purpose open-source software reveals the unexplored difference between SATD manifestation in these distinct contexts. 
With SSW containing 9.25x more Scientific Debt than general-purpose open-source software, it is probable that SSW, due to its specialized nature and the unique challenges faced by domain experts, exhibits this higher prevalence of Scientific Debt compared to general-purpose open-source software. 

Furthermore, we noticed a difference in the distribution of SATD between the two types of software. We observed that SSW contains 4.93x more SATD than general-purpose open-source software and that general-purpose open-source software contains 3.09x more documentation debt than SSW. \end{tcolorbox}

\section{Discussion}
\label{sec:discussion}
In this section, we discuss limitations, how this study helps understand and address SATD, and how it can help maintain the integrity and efficiency of SSW over time.

\subsection{Limitations}

Despite the valuable insights provided by this study, several limitations should be considered:

\begin{enumerate} 
    \item \textbf{Dataset Limitations:} The dataset used in this study is limited to specific repositories, which may not fully represent all SSW domains during training. The selection criteria for evaluated repositories in RQ2 may also introduce bias. Furthermore, the dataset and evaluated repositories are all derived from publicly available sources, which may not capture all instances of SATD present in proprietary or private repositories.
    
    \item \textbf{Labeling and Annotation Issues:} The accuracy of the SATD classification depends heavily on the quality of labeling and annotation. Any errors or inconsistencies in the labeling process during training could impact the reliability of the results. Additionally, the granularity of debt types may vary, leading to potential misclassification and affecting the overall analysis.

    \item \textbf{Temporal and Contextual Changes:} The study examines SATD at a specific point in time, and the results may not account for changes over time or evolving practices in software development. Future studies would benefit from longitudinal analysis to assess how SATD trends shift with new technologies and methodologies.
    
    \item \textbf{Generalizability:} While the study provides a detailed analysis of SATD in selected repositories, the generalizability of the results to all SSW or broader contexts is limited. Further research may be needed to validate the findings across a wider range of repositories and software types to provide a more comprehensive understanding of SATD.
\end{enumerate}

\subsection{Research Question Analysis and Implications}
\label{sec:Research Question Analysis and Implications}
Through conducting extensive testing of many untested fine-tuned models in the application of identifying SATD, we introduce a new best performing model for SATD identification and classification. Our testing results of these models show that conducting a thorough grid search to identify optimal hyperparameters, and dataset balancing both increase model performance. Furthermore, through testing models larger than 3B parameters, we are the first to identify that despite these models being much larger than smaller transformers such as BERT-large, they don't outperform smaller models across all metrics.

Our comparison of SATD between SSW and general-purpose open-source software offers novel insights into how TD manifests in these contexts. The 9.25x higher prevalence of Scientific Debt in SSW underscores the unique challenges domain experts face to translate complex scientific theories into computational models. Furthermore, our findings show that SSW contains 4.92x SATD. These findings imply that SATD in SSW is often closely tied to the specialized nature of the domain, and that SATD management in SSW is lacking compared to the highly studied general-purpose open-source software.

\subsection{Role of Domain Knowledge in Shaping SATD Across Software Types}
Our findings from RQ2 reveal a significant difference in the quantity of SATD present in SSW versus general-purpose open-source software, a pattern that can be understood through the knowledge acquisition model proposed by Kelly~\cite{Kelly2015ScientificSD}. Kelly views SSW development as inherently driven by the need to acquire and apply domain-specific knowledge, where software serves as a tool for scientific discovery rather than an end in itself. This knowledge-driven approach helps explain why our results show that SSW exhibits a higher prevalence of Scientific Debt, as the software is deeply intertwined with complex, evolving scientific theories and models. 
In contrast, general-purpose open-source software with a much smaller quantity of SATD and Scientific Debt suggests a focus---more on software structure and design than on the specific domain being addressed---a pattern supported by the findings from a large-scale industry survey indicating that structured development practices in larger systems correlate with lower technical debt prevalence \cite{ramavc2022prevalence}.

This contrast in the quantity of SATD reflects the distinct priorities of developers in these two contexts. In SSW, the focus is often on capturing and representing the scientific domain accurately, leading scientists to identify and document issues related to scientific concepts, even if other areas like code quality or design are not fully addressed, which we observe through the differing quantities of documentation debt in the two domains. This could be explained by the fact that scientists, as observed by Carver et al.~\cite{carver2007software}, Pinto et al.~\cite{Pinto2018HowDS}, and Hannay et al.~\cite{hannay2009scientists}, often lack formal education in software engineering and tend to learn programming through self-study or on the job.

Our findings align with the knowledge model proposed by Kelly and suggests that the type of SATD present in a given software project is strongly influenced by the domain expertise of its developers. In SSW, SATD is more prevalent because domain scientists, lacking formal software engineering training, prioritize advancing scientific progress, which can lead to accumulated inaccuracies and assumptions over time rather than an emphasis on code quality management. On the other hand, general-purpose software developers, with a stronger focus on engineering best practices, tend to document and address SATD. This highlights the need for a more nuanced approach to SATD management, which considers the software's purpose and the background and expertise of the developers involved.

\section{Conclusion}
\label{sec:conclusion}

In this study, using a transformer-based SATD identification and classification tool fine-tuned toward identifying the intricacies of SSW, we have identified and analyzed the differences in SATD between open-source SSW and general-purpose open-source software. 
Our approach leverages state-of-the-art transformer-based models, to accurately classify various types of SATD in scientific repositories. The models' F1-scores outperformed previous literature, demonstrating their effectiveness in identifying SATD and distinguishing between different types of TD.

The findings of this study provide valuable insights into SATD in SSW. By highlighting patterns and tendencies in SATD, our work opens up new avenues for research on TD in SSW development. Furthermore, the labeled dataset created in this study can serve as a benchmark for future work in the field, facilitating comparative studies and advancements in SATD detection techniques.

While this research represents a significant step forward in understanding and managing SATD in SSW, several areas remain for future exploration. Enhancing the datasets' diversity and size, including contextual information and sentiment, leveraging multiple sources, and conducting longitudinal studies to observe how SATD evolves in scientific projects are promising directions for future work. Additionally, potentially integrating our classifier with existing software development workflows could help developers proactively manage and reduce SATD, ultimately leading to more sustainable and maintainable SSW.

This study offers a novel analysis of SATD in SSW, paving the way for future research and innovation in this field. By addressing SATD, our work aims to elevate software quality and promote long-term sustainability in scientific computing, ultimately empowering more reliable and impactful research outcomes.

\bibliographystyle{plain}

\bibliography{bibliography}

\end{document}